\documentclass[preprint,preprintnumbers,amsmath,amssymb]{revtex4}
\usepackage{bm}
\usepackage{newtxmath,newtxtext}
\usepackage{graphicx}
\usepackage{textcomp}
\usepackage{float}
\usepackage{booktabs}
\usepackage{dcolumn,enumerate}
\usepackage{ragged2e}
\usepackage{hyperref}
\hypersetup{colorlinks,citecolor=blue}
\hypersetup{colorlinks=true, linkcolor=blue, filecolor=blue, urlcolor=blue}

\usepackage{xcolor}
\usepackage{epsfig}
\usepackage{caption}
\usepackage{appendix}
\usepackage{subcaption}
\usepackage{graphicx} 
\usepackage{url}
\begin{document}

\title{\textbf{Revisiting the Penrose Process in Rotating Black Holes with Quantum Corrections: Implications for  Energy Extraction and Irreducible Mass}}

\author{\textsuperscript{1}Urooj Fatima}
\email{wafa4494@gmail.com}

\author{\textsuperscript{1}G. Abbas}
\email{ghulamabbas@iub.edu.pk}

\affiliation{\textsuperscript{1}Department of Mathematics, The Islamia University of Bahawalpur, Bahawalpur-63100, Pakistan}

\date{}

\begin{abstract}
We explore the extraction of energy from a rotating black hole spacetime modified by a quantum correction parameter \( \alpha \). Focusing on particle splitting within the ergoregion, we analyze the Penrose process and compute the extraction efficiency \( \eta \) as a function of both the spin parameter \( a \) and the quantum correction parameter \( \alpha \). Our results show that increasing \( \alpha \) induces an inward shift of the event horizon and the static limit, resulting in a modest expansion of the ergoregion. This geometric change significantly enhances the energy extraction potential. By numerically solving the horizon equation, we determine a maximum extraction efficiency of 11.64\%. Additionally, we derive the expression for the irreducible mass, highlighting its fundamental role in constraining the amount of extractable rotational energy. Overall, our findings demonstrate that quantum corrections have a substantial impact on black hole energetics, leading to marked deviations from predictions based on classical Kerr theory.

\end{abstract}

\maketitle
\section{Introduction}
The Kerr solution reveals that rotating black holes host unique spacetime regions, such as the ergosphere, that facilitate the extraction of their immense rotational energy via processes like the Penrose mechanism \cite{wikipedia-penrose, kottke2018harvest}. However, quantum gravity corrections can fundamentally alter the properties governing energy extraction efficiency, it becomes vital to investigate their impact on these black holes. This study is therefore dedicated to investigating precisely how these quantum corrections alter the energy extraction capabilities of rotating black holes. Over the past few years, observational discoveries, such as the Event Horizon Telescope (EHT) imaging of M87 and Sgr A* \cite{ali2025shadows}, have begun to place empirical constraints on black hole properties, including spin and possible deviations from the classical Kerr solution. These considerations are an incentive to consider more general black hole spacetimes, such as ones with quantum corrections, since even small deformations of the event horizon or ergoregion can lead to observable astrophysical effects.

The ergoregion is what makes rotating black holes special: this is the region outside the event horizon where the frame-dragging force is so strong that even the fastest observers cannot remain stationary concerning distant sources \cite{ma2025distinguishing}. The result of being in this ergoregion is that the Penrose process \cite{penrose1969gravitational} can occur, where particles can split, with one portion falling with more negative energy into the black hole, and the second fragment escaping with more energy than it had initially. Many scientists have used classical general relativity to explore how rotational energy is removed from black holes. When a Kerr black hole has been spun to the extremal limit, the Penrose process can have the most excellent efficiency, approximately 20.7 percent \cite{bhat1985energetics}. Whether one is discussing astrophysics or general relativity, the study is essential because rotating black holes have a direct impact on high-energy astrophysical conditions. Particularly, adjusting gravity or quantum-correcting models might offer an indirect detection of new physics beyond general relativity, as they could influence energy extraction efficiencies and reveal variations in jet power and emission \cite{tursunov2019fifty}, which would be considered notable.

Afterwards, researchers investigated how energy may be extracted from more complex black holes, including those with charge, such as Kerr-Newman \cite{bhat1985energetics}, and those with magnetic fields \cite{chakraborty2024magnetic} that affect them. Often, having an electric charge or external fields lowers the area of the ergoregion and makes energy extraction less efficient \cite{bhat1985energetics}. Alternatively, theories in gravity and models in higher dimensions, which include braneworld scenarios, have revealed that parameters such as tidal charge can extend the ergoregion and increase efficiency. As demonstrated \cite{khan2019particle} that a greater tidal charge, along with a rapid spin, enhances the ergosphere, thereby expanding the device's potential for energy extraction efficiency. Deformed black holes (e.g., higher-dimensional non-Kerr spacetimes\cite{ghosh2014higher}), scalar hair, and external magnetic fields  in magnetized Reissner–Nordström geometries \cite{shaymatov2022efficiency} or regular black hole mimickers \cite{viththani2024magnetic} have been studied; results show that changes in spacetime geometry are crucial for the Penrose process.

The impact of quantum gravity corrections on black holes is profound, not only regularising singularities but also dramatically altering their fundamental characteristics, including causal structure, horizon geometry, and thermodynamics \cite{modesto2004disappearance, bonanno2000renormalization, ali2009discreteness}. Emerging research further demonstrates that these quantum effects significantly influence the innermost stable circular orbit (ISCO) and the ergoregion \cite{mustafa2025testing} – areas critical for energy extraction in rotating black holes. While the thermodynamics of quantum-corrected black holes has been a subject of considerable study \cite{liu2025quantum, blanchette2021black, ongole2024revisiting}, our knowledge of their energy and dynamical properties remains remarkably limited. This is especially true for rotating black holes, where the ergoregion plays a paramount role. This pressing need to understand these unexplored aspects forms the core motivation of the present work.

Our work examines the extraction of energy from a rotating quantum-corrected black hole RQCBH that follows a metric recently derived using a modified Newman-Janis algorithm on a static quantum-corrected solution \cite{ali2025shadows}. As a result, the metric changes from the classical Kerr solution to the quantum-corrected regime in a smooth way, with the parameter $\alpha$ linking the two domains. Unlike previous studies, our work differs in that we incorporate quantum gravity corrections directly into the geometry of the rotating black hole, thereby preserving important symmetries and enabling a systematic analysis of their physical implications. Internally, the ergoregion of the Penrose process judiciously relies on the presence and the properties of negative energy states, so any changes in the shape, size, or position of the ergosphere caused by quantum effects can significantly impact the ability to utilise the Penrose process for energy extraction. We study the motion of neutral particles in the quantum-corrected spacetime, determine the criteria for negative energy orbits, and compute the efficiency of energy extraction as functions of both the spin parameter a and the quantum correction parameter $\alpha$. To examine whether quantum effects enhance or hinder energy extraction from a black hole, we compare our results with those for classical and braneworld Kerr black holes.

Similarly, research made possible by gravitational wave astronomy and instruments such as LIGO, Virgo, and the future LISA can now be used to investigate quantum-corrected black holes \cite{giddings2016gravitational, barausse2020prospects}. Altering the values of the ISCO, the orbital angular momentum, and energy extraction efficiency can modify the gravitational wave signals of inspiraling binaries \cite{zi2025eccentric}, allowing us to constrain parameters of quantum gravity independently.

This paper is structured into distinct sections to explore the rotating quantum-corrected black hole systematically. \textbf{Section} \ref{s2} introduces the black hole metric and examines its key geometric features, including the event horizon, static limit, and ergoregion. Following this, \textbf{Section} \ref{s3} derives the equations governing the motion of neutral particles, analysing their behaviour and the angular velocity of the particles. The discussion then moves to \textbf{Section} \ref{s4}, which focuses on the formation of negative energy orbits, including a discussion of the Wald inequality and an evaluation of the Penrose process's efficiency. In \textbf{Section} \ref{s5}, the concept of irreducible mass is explored in detail. Finally, \textbf{Section} \ref{s6} presents a summary of the main results.

\section{Rotating Quantum Corrected Black Hole } \label{s2}

In this paper, we use a modified Newman–Janis algorithm to transform a loop quantum gravity-inspired static black hole into a rotating quantum-corrected black hole (RQCBH) geometry. The result is a stationary and axisymmetric spacetime that is a generalised Kerr metric, obtained by introducing a quantum correction parameter, denoted as $\alpha$. The line element of the RQCBH \cite{ali2025shadows} in Boyer–Lindquist coordinates (t, r, $\theta$, $\phi$) is

\begin{equation}
\begin{split}
ds^2 = & -\left( \frac{\Delta - a^2 \sin^2\theta}{\Sigma} \right) dt^2 
- 2a \sin^2\theta \left( 1 - \frac{\Delta - a^2 \sin^2\theta}{\Sigma} \right) dt\,d\phi \\
& + \sin^2\theta \left[ \Sigma + a^2 \sin^2\theta \left( 2 - \frac{\Delta - a^2 \sin^2\theta}{\Sigma} \right) \right] d\phi^2 \\
& + \frac{\Sigma}{\Delta} dr^2 + \Sigma\, d\theta^2, \label{a1}
\end{split} 
\end{equation}

where
\begin{equation}
\Delta = r^2 + a^2 - 2 M r + \frac{\alpha M^2}{r^2}, 
\quad 
\Sigma = r^2 + a^2 \cos^2\theta.
\end{equation}

Here M represents the ADM mass of the RQCBH, which, in asymptotically flat spacetime, is equivalent to the total gravitational mass as determined by a remote observer, a is the rotation parameter, and $\alpha$ is the correction parameter that measures the potential deviation of metric (\ref{a1}). The term $\frac{\alpha M^2}{r^2}$
 is introduced by effective quantum gravity and changes the shape of the region close to the horizon. At extensive radial distances, this correction becomes insignificant, maintaining the asymptotic configuration of the Kerr spacetime. When $\alpha$ is set to zero, the metric becomes the classical Kerr solution, which agrees with general relativity in the absence of quantum effects. 
 
\subsection{Horizon Structure for RQCBH} 

\begin{figure}
\includegraphics[width=0.7\linewidth, height=2.8in]{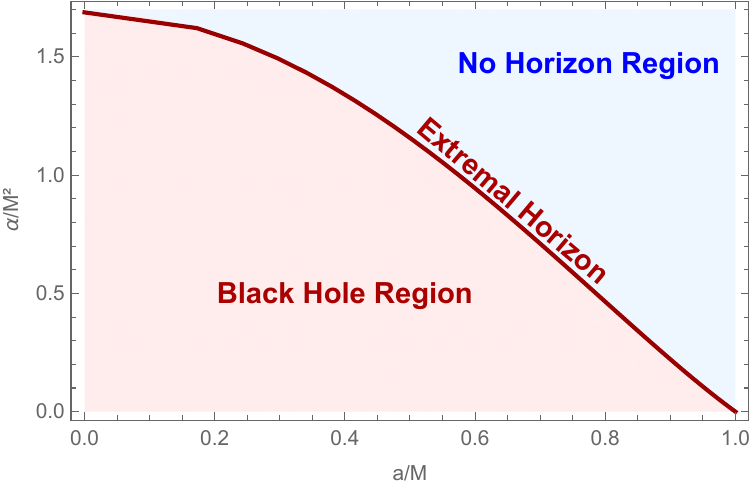}
\caption{Parameter space of the rotating quantum-corrected black hole (RQCBH) as a function of spin a and quantum correction parameter $\alpha$.} \label{Fig: a}
\end{figure}

The parameter space of the rotating quantum-corrected black hole (RQCBH) is illustrated in \textbf{Fig.}~\ref{Fig: a}, which maps the spin parameter \( a \) against the quantum correction parameter \( \alpha \). This space is divided into three distinct regions. The \textbf{black hole region} includes combinations of \( a \) and \( \alpha \) for which spacetime admits two horizons, event and Cauchy, indicating physically viable black hole solutions. The dark red curve marks the \textbf{ extremal boundary}, where the two horizons coincide. Beyond this curve, no event horizon exists, signaling either a naked singularity or a horizonless spacetime. This boundary highlights how quantum corrections and rotation jointly determine the conditions for black hole formation.

As \( \alpha \) increases, the extremal value of the spin parameter \( a_E \) decreases, and vice versa—larger \( a \) values reduce the allowed range of \( \alpha \). Thus, \textbf{Fig.}~\ref{Fig: a} serves as a critical reference for identifying parameter ranges that yield regular black hole solutions. Restricting our analysis to this region ensures the physical relevance of the results.

 The figure also illustrates how the locations of the event horizon (\( r_+ \)) and Cauchy horizon (\( r_- \)) evolve with varying \( \alpha \) and \( a \). As \( \alpha \) increases, the horizons approach one another and eventually merge at extremality, beyond which no black hole exists.

\begin{figure}[H]
\includegraphics[width=0.7\linewidth, height=2.8in]{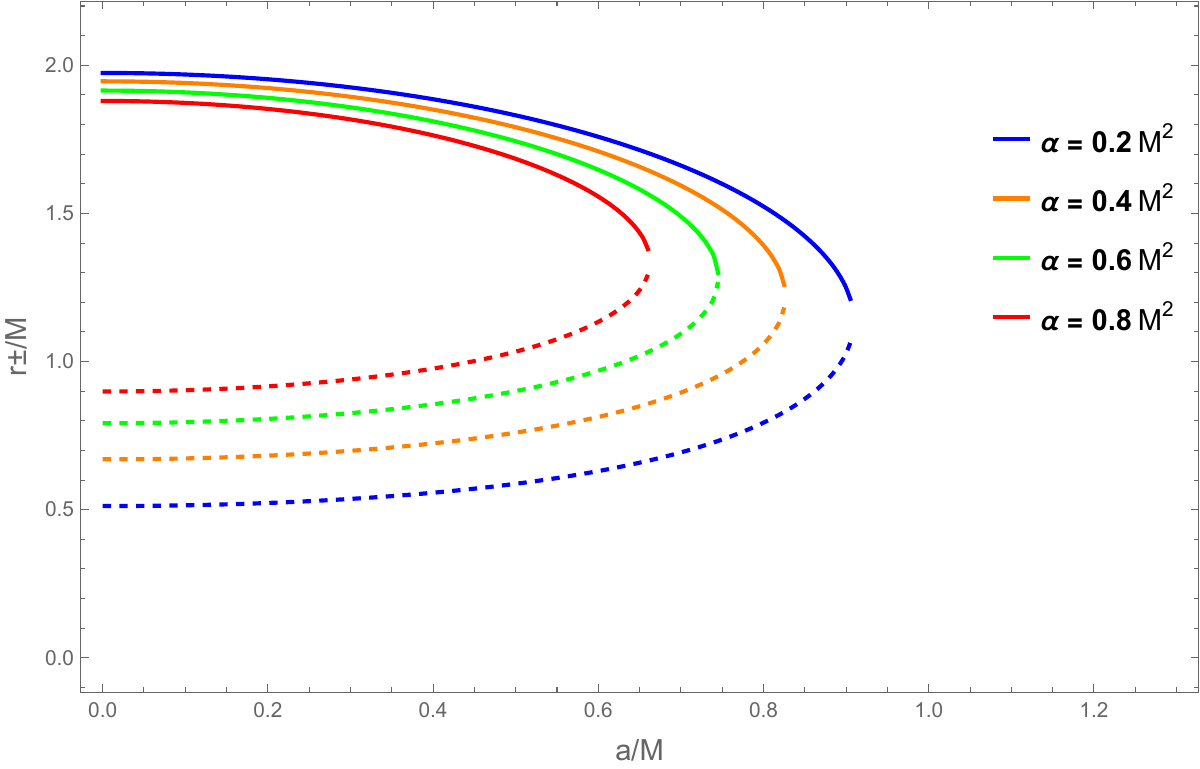}
\caption{ Solid and dotted curves are representing Event and Cauchy Horizons repectively for RQCBH.}\label{Fig: b}
\end{figure}

\begin{figure}[htbp]
    \centering

    \makebox[\textwidth][c]{
        (a)\includegraphics[width=0.45\textwidth]{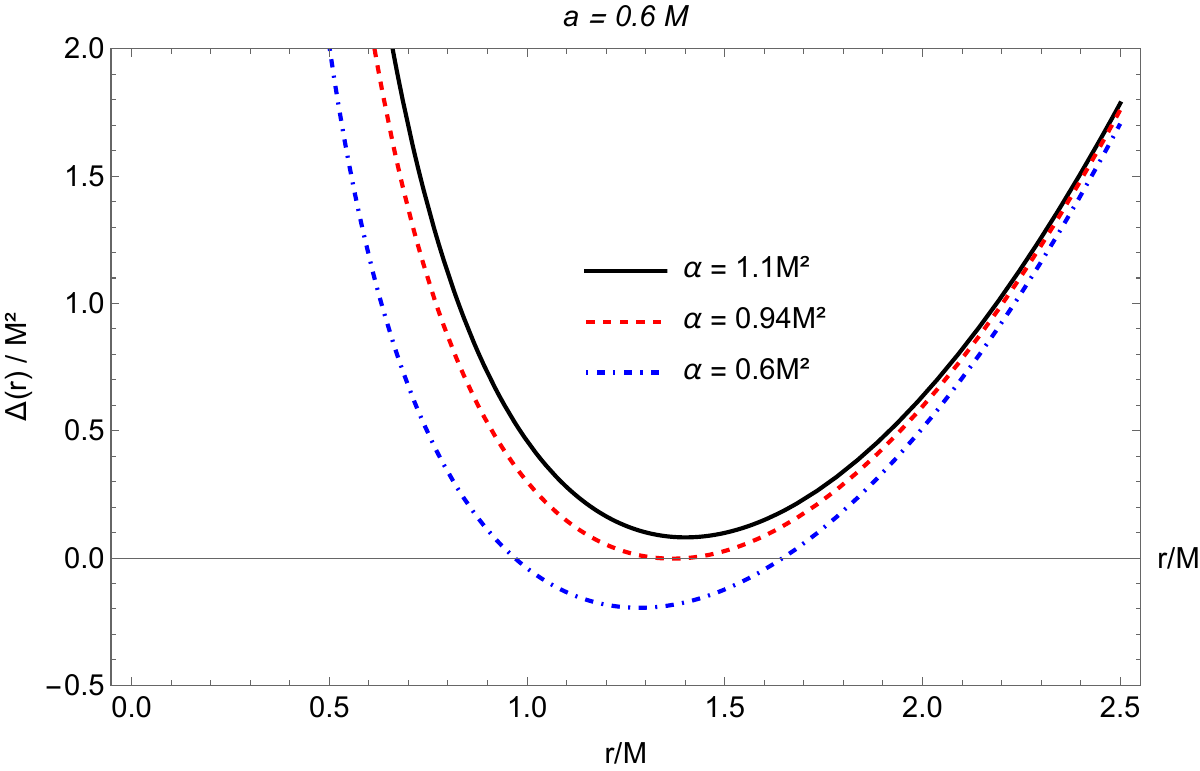}
        \hspace{2em}
        (b)\includegraphics[width=0.45\textwidth]{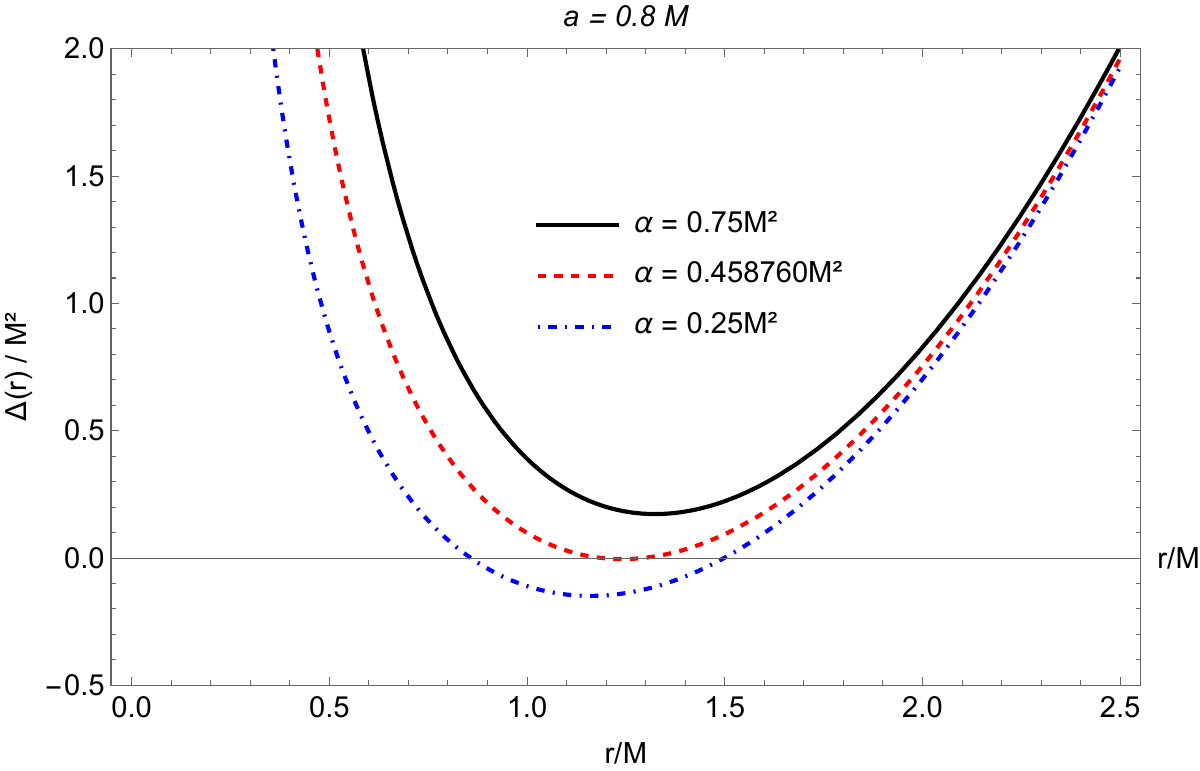} 
    }

    \vspace{2em}

    \makebox[\textwidth][c]{
      (c)  \includegraphics[width=0.45\textwidth]{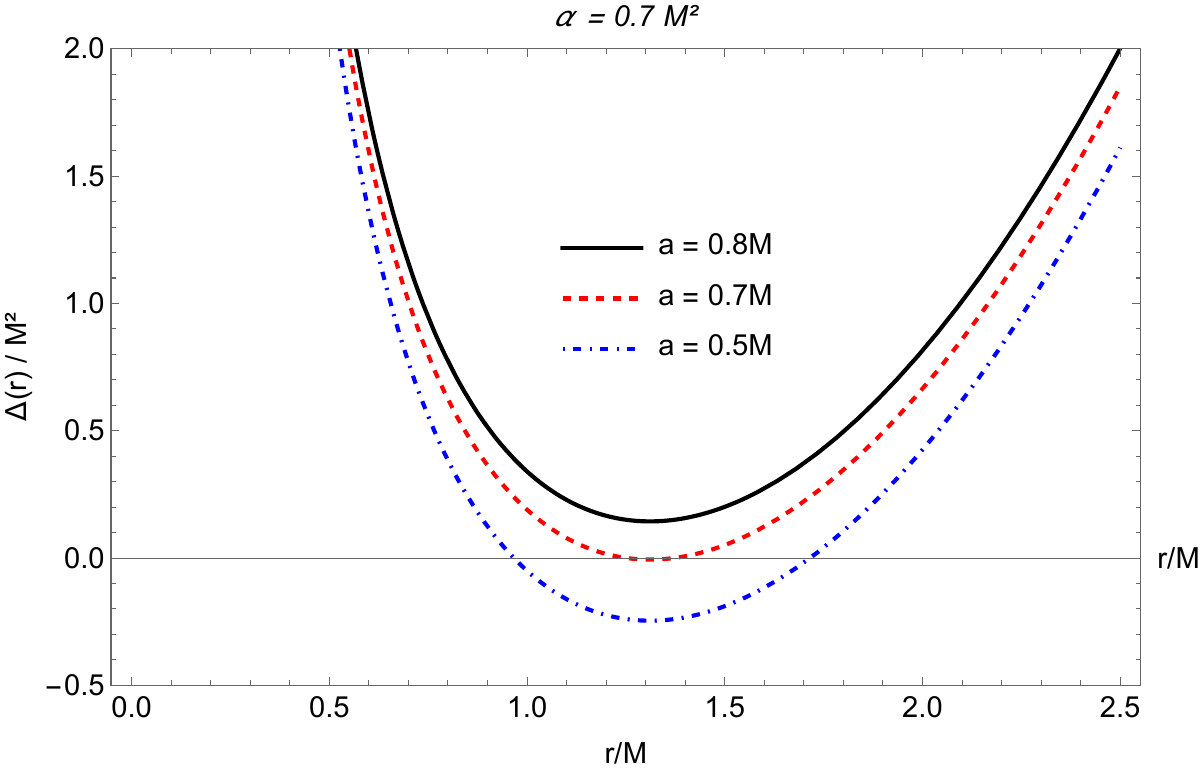}
        \hspace{2em}
        (d)\includegraphics[width=0.45\textwidth]{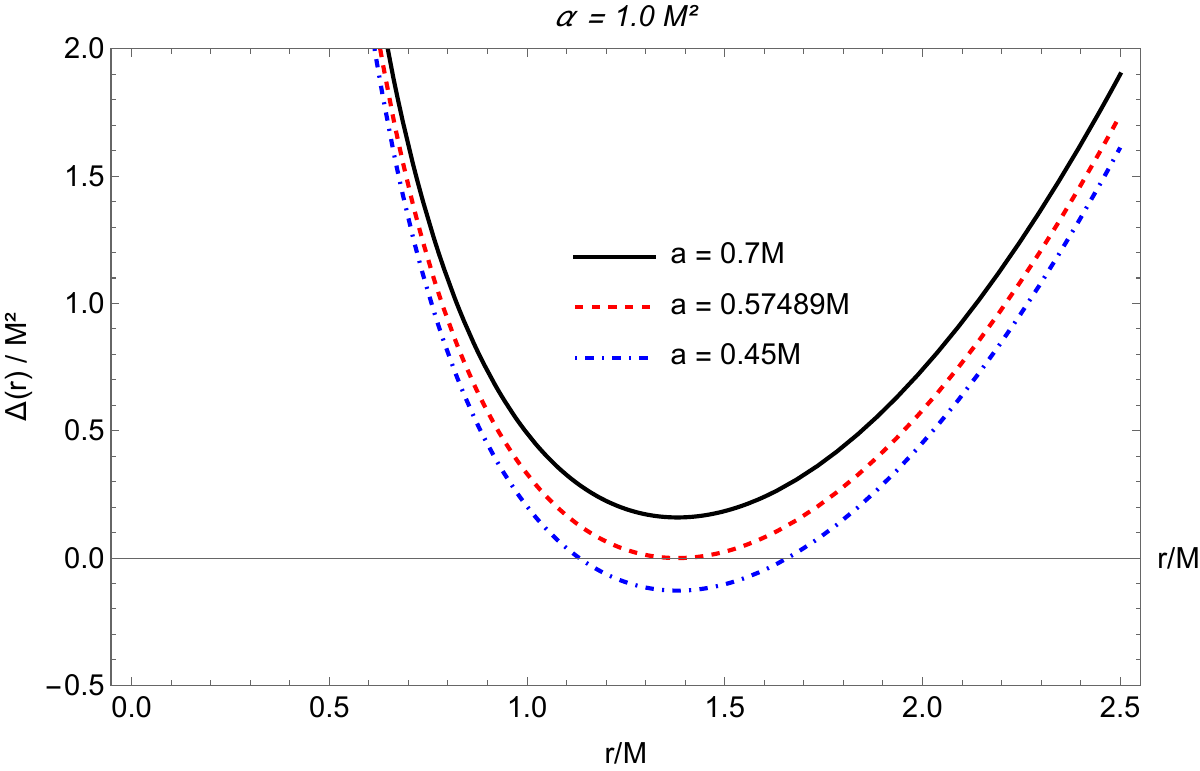} 
    }

    \caption{Illustrates the behavior of the metric function $\Delta(r)$ as a function of $r$ for different values of $a$ and $\alpha$. The roots of $\Delta(r) = 0$ correspond to the positions of the Cauchy and event horizons in a rotating quantum-corrected black hole (RQCBH). }
    \label{Fig: c}
\end{figure}
The upper plots in \textbf{Fig.} [\ref{Fig: c}] demonstrate that as $\alpha$ increases for a fixed $a$, the horizon radii shrink. Beyond the extremal value, i.e., when $\alpha > \alpha_E$, no real roots exist, indicating the absence of horizons. Similarly, the lower panels show that as $a$ increases for a fixed $\alpha$, the horizons draw closer together, eventually merging at extremality or disappearing entirely. The red lines represent scenarios where the two horizons coalesce.

\begin{table}[h!]
\centering
\caption{The ergoregion ($\delta = r_{\text{es}}^+ - r_+$), event horizon ($r_+$) and static limit ($r_{\text{es}}^+$) for different $a$ and $\alpha$ values.}
\begin{tabular}{|c|ccc|ccc|ccc|ccc|}
\hline
\multicolumn{1}{|c|}{$\alpha$} & 
\multicolumn{3}{c|}{$a = 0.2$} & 
\multicolumn{3}{c|}{$a = 0.3$} & 
\multicolumn{3}{c|}{$a = 0.4$} & 
\multicolumn{3}{c|}{$a = 0.5$} \\
\hline
 & $r_+$ & $r_{\text{es}}^+$ & $\delta$ 
 & $r_+$ & $r_{\text{es}}^+$ & $\delta$ 
 & $r_+$ & $r_{\text{es}}^+$ & $\delta$ 
 & $r_+$ & $r_{\text{es}}^+$ & $\delta$ \\
\hline
0.0 & 1.9798 & 1.9899 & 0.0102 & 1.9539 & 1.9772 & 0.0233 & 1.9165 & 1.9592 & 0.0427 & 1.8660 & 1.9354 & 0.0694 \\
0.1 & 1.9665 & 1.9769 & 0.0104 & 1.9399 & 1.9639 & 0.0240 & 1.9013 & 1.9453 & 0.0440 & 1.8490 & 1.9208 & 0.0718 \\
0.2 & 1.9527 & 1.9634 & 0.0107 & 1.9252 & 1.9499 & 0.0247 & 1.8853 & 1.9308 & 0.0455 & 1.8309 & 1.9855 & 0.0746 \\
0.3 & 1.9382 & 1.9492 & 0.0111 & 1.9098 & 1.9354 & 0.0256 & 1.8684 & 1.9156 & 0.0472 & 1.8115 & 1.8894 & 0.0778 \\
0.4 & 1.9229 & 1.9344 & 0.0115 & 1.8936 & 1.9201 & 0.0265 & 1.8504 & 1.8995 & 0.0491 & 1.7907 & 1.8723 & 0.0816 \\
0.5 & 1.9069 & 1.9188 & 0.0119 & 1.8763 & 1.9039 & 0.0276 & 1.8312 & 1.8826 & 0.0514 & 1.7682 & 1.8541 & 0.0860 \\
0.6 & 1.8900 & 1.9023 & 0.0124 & 1.8580 & 1.8868 & 0.0288 & 1.8111 & 1.8645 & 0.0534 & 1.7434 & 1.8347 & 0.0914 \\
0.7 & 1.8719 & 1.8849 & 0.0129 & 1.8384 & 1.8686 & 0.0303 & 1.7881 & 1.8452 & 0.0572 & 1.7157 & 1.8138 & 0.0981 \\
0.8 & 1.8526 & 1.8662 & 0.0136 & 1.8171 & 1.8491 & 0.0320 & 1.7634 & 1.8244 & 0.0610 & 1.6840 & 1.7990 & 0.1069 \\
0.9 & 1.8317 & 1.8461 & 0.0144 & 1.7939 & 1.8281 & 0.0341 & 1.7357 & 1.8017 & 0.0660 & 1.6466 & 1.7657 & 0.1192 \\
1.0 & 1.8090 & 1.8243 & 0.0154 & 1.7682 & 1.8050 & 0.0368 & 1.7040 & 1.7776 & 0.0727 & 1.5991 & 1.7374 & 0.1382 \\
\hline
\end{tabular}
\label{T1}
\end{table}

\textbf{Table} [\ref{T1}] provides a detailed analysis of the ergoregion structure, illustrating how the event horizon radius ($r_+$), the static limit surface ($r_{es+}$), and the resulting ergoregion width ($r_{es+} - r_+$) vary with different values of the spin parameter $a$ and the quantum correction parameter $\alpha$.

The results indicate that an increase in $\alpha$ leads to a decrease in both $r_+$ and $r_{es+}$. However, the ergoregion width increases, revealing an expanding region where frame-dragging effects dominate. The dependence on $a$ further highlights the role of black hole rotation in enlarging the ergoregion. These trends collectively emphasize the combined influence of spin and quantum corrections on the size and structure of the ergoregion.

\begin{figure}[htbp]
    \centering

    \includegraphics[width=0.23\textwidth]{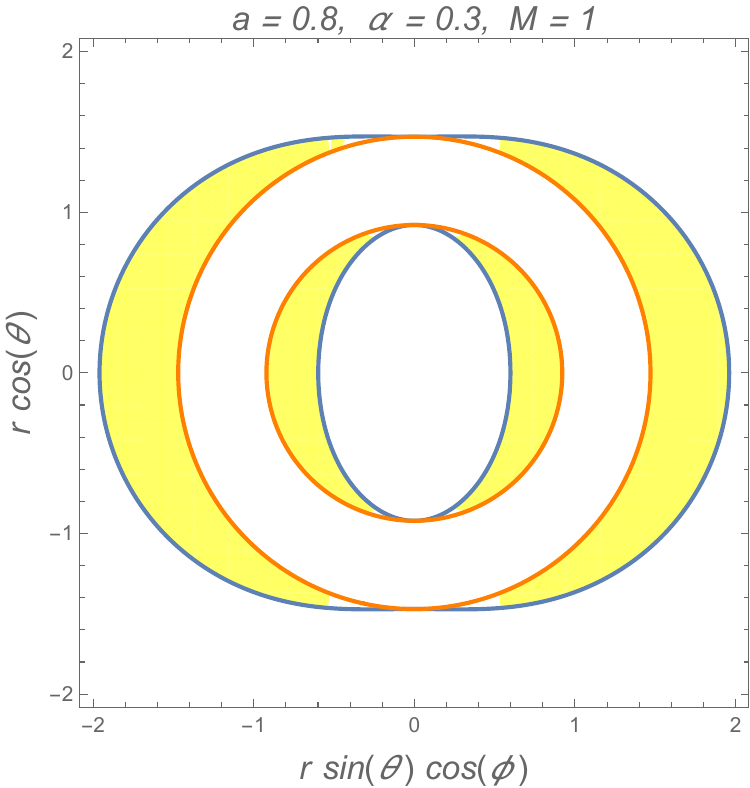}
    \hfill
    \includegraphics[width=0.23\textwidth]{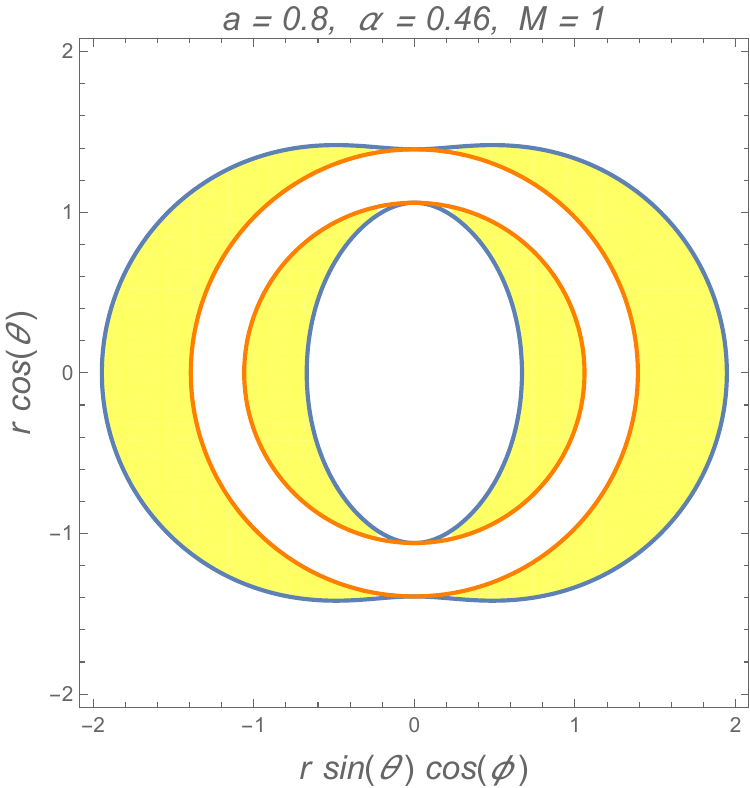}
    \hfill
    \includegraphics[width=0.23\textwidth]{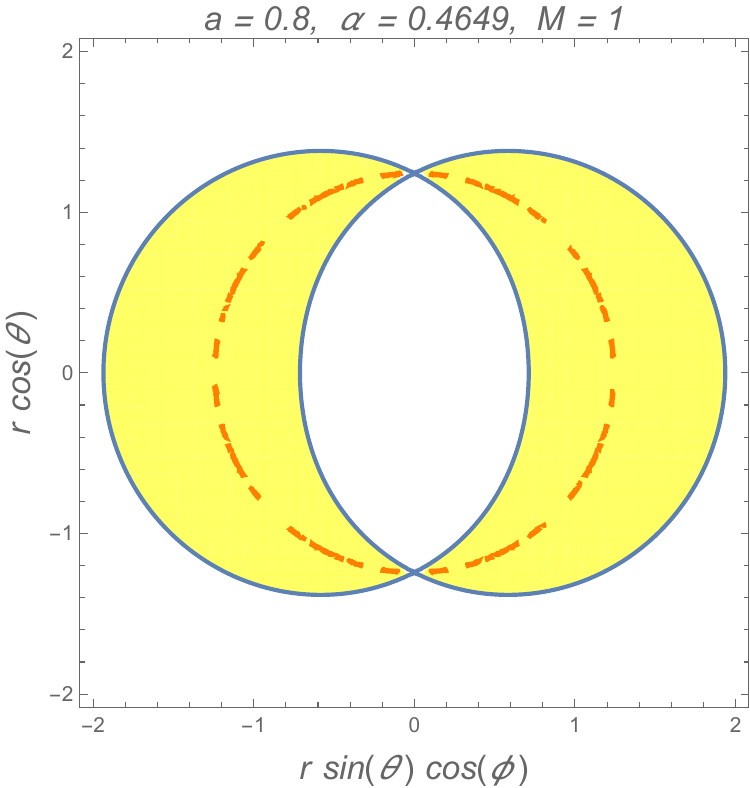}
    \hfill
    \includegraphics[width=0.23\textwidth]{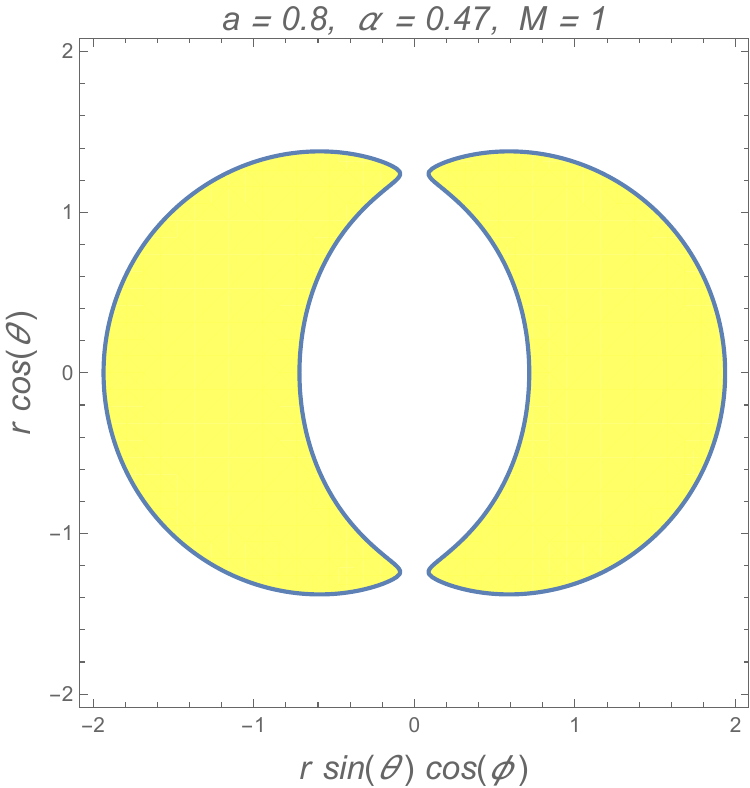}

    \vspace{2em}

    \includegraphics[width=0.23\textwidth]{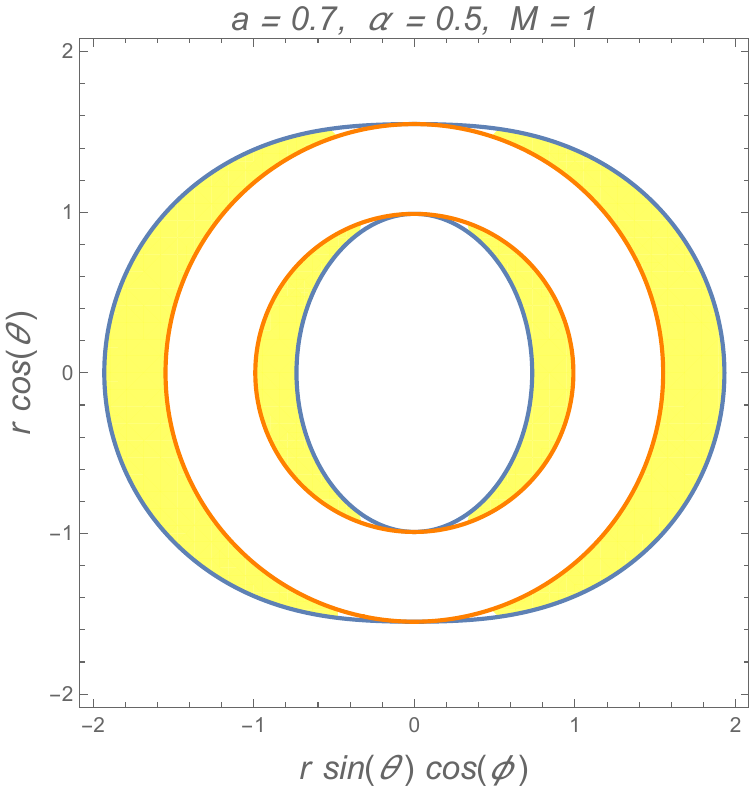}
    \hfill
    \includegraphics[width=0.23\textwidth]{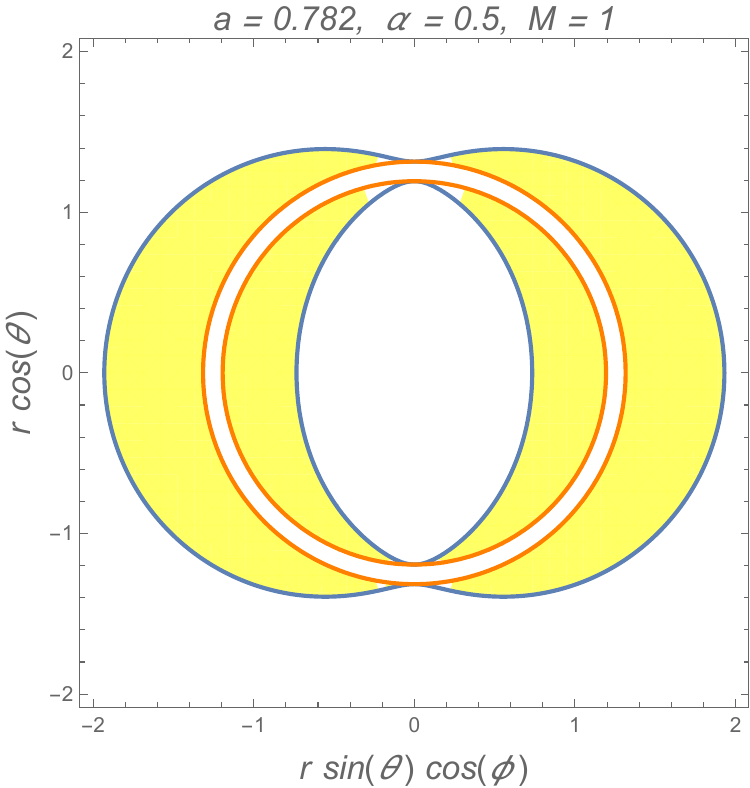}
    \hfill
    \includegraphics[width=0.23\textwidth]{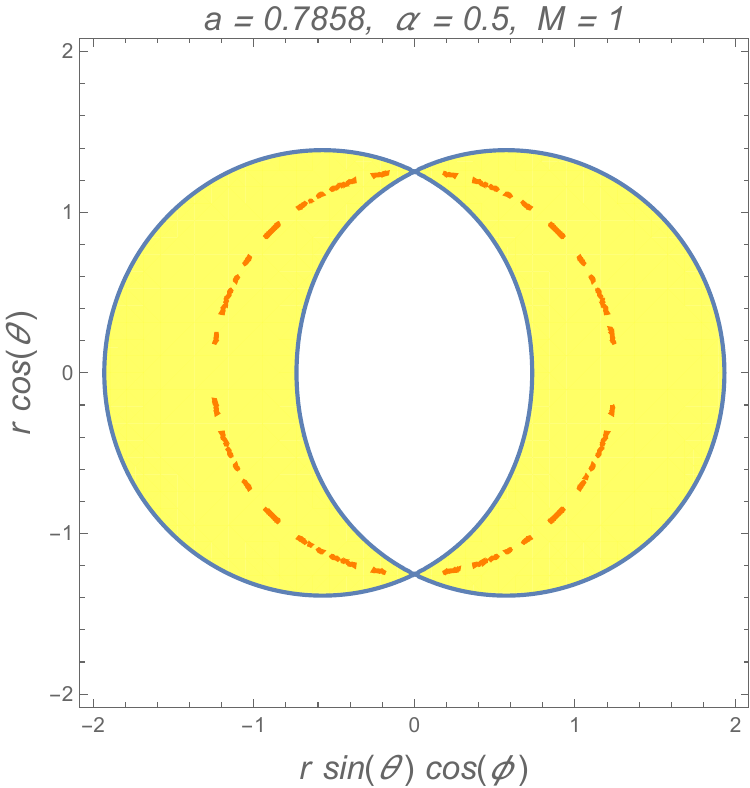}
    \hfill
    \includegraphics[width=0.23\textwidth]{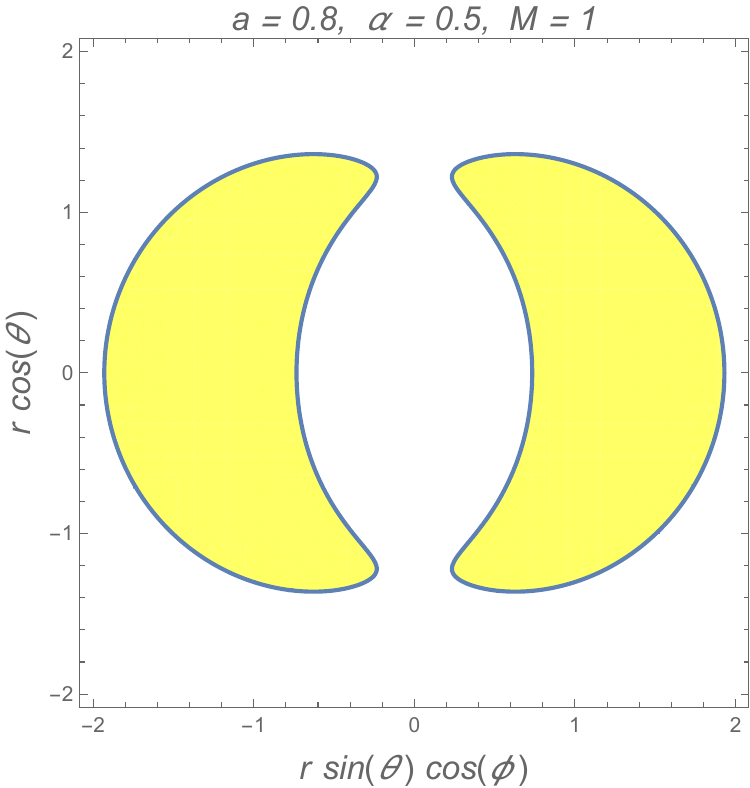} 

    \caption{cross-sectional view of the rotating quantum-corrected black hole, illustrating the event horizon (outer red curve), the stationary limit surface (outer blue curve), and the ergoregion for various values of the quantum correction parameter $\alpha$.}
    \label{fig:4}
\end{figure}

\textbf{Fig.} [\ref{fig:4}] visualizes the ergoregion structure corresponding to the tabulated values in \textbf{Table} [\ref{T1}]. The plots clearly demonstrate that the ergoregion increases in size as either the spin parameter $a$ or the quantum correction parameter $\alpha$ increases. This expansion signifies an enhancement of frame-dragging effects due to both rotation and quantum corrections. Larger values of $a$ and $\alpha$ lead to a more pronounced and extensive ergoregion. For sufficiently large parameter values, the ergoregion manifests as multiple disconnected regions forming between the event horizon and the static limit surface. This behavior underscores the significant and combined role of spin and quantum effects in shaping the ergoregion.

\section{Particle Dynamics in the Background of RQCBH} \label{s3}

To examine the motion of particles orbiting around a rotating black hole, we shall first look at the general motion of neutral particles in the gravitational field produced by the black hole. To simplify and clarify the discussion, we consider motion in the equatorial plane, which is typically the most physically meaningful case in astrophysical applications. The geometry of spacetime in the black hole dictates the path of particles. The metric of this geometry is a mathematical description of how distances and times are measured close to the black hole.

We apply the Lagrangian formalism to derive the equations of particle motion. The metric is used to form the Lagrangian of a test particle in the specified spacetime, and the equations of motion (geodesics) are found by using the Euler-Lagrange equations. These are equations that detail the development of the position and velocity of the particle relative to its proper time.

The Lagrangian for a test particle of unit mass is given by
\begin{equation}
\mathcal{L} = \frac{1}{2} g_{\mu\nu} \dot{x}^\mu \dot{x}^\nu, \label{a11}
\end{equation}
From above equation, we obtain the generalized momenta corresponding to the cyclic coordinates
\begin{align}
-p_t &= g_{tt} \dot{t} + g_{t\phi} \dot{\phi} = E, \label{a2} \\
p_\phi &= g_{t\phi} \dot{t} + g_{\phi\phi} \dot{\phi} = L \label{a3}.
\end{align}

Solving the above system of \textbf{Eqns.} (\ref{a2}) and (\ref{a3}) for \(\dot{t}\) and \(\dot{\phi}\), we find:
\begin{align}
\dot{t} &= \frac{1}{r^2 \Delta} \left[ E a^2 (\frac{\alpha M^2}{r^2} - r(2M + r)) + a L (\frac{\alpha M^2}{r^2} - 2Mr) - E r^4 \right], \label{a4} \\
\dot{\phi} &= \frac{1}{r^2 \Delta} \left[ a E (\frac{\alpha M^2}{r^2} - 2Mr) + L (\frac{\alpha M^2}{r^2} + r(r - 2M)) \right] \label{a5}.
\end{align}

Expressions (\ref{a4}) and (\ref{a5}) are derived from the structure of the rotating black hole metric (\ref{a1}) and are model-dependent; for the RQCBH, appropriate replacements of the metric functions (such as \(\Delta(r)\) should be applied here.

To study the dynamics of the particles in the spacetime of the  RQCBH further, it is usually beneficial to switch from the Lagrangian to the Hamiltonian formulation of the dynamics. The Hamiltonian method offers a coherent and intuitive method of formulating geodesic motion in terms of the coordinates of the particle and its conjugate momenta. For a test particle of unit mass, the Hamiltonian in a curved spacetime is defined as:

\begin{equation*}
\mathcal{H} = \frac{1}{2} g^{\mu\nu} p_\mu p_\nu,
\end{equation*}
where \(g^{\mu\nu}\) are the components of the inverse metric and \(p_\mu\) are the covariant momenta. This expression represents the total energy of the particle in terms of the spacetime geometry and its motion. Replacing the expressions of the momenta and imposing the normalisation condition on the timelike geodesics, one gets the radial equation of motion. 

\begin{equation}
2\mathcal{H} = E \dot{t} + L \dot{\phi} + \frac{r^2}{\Delta} \dot{r}^2 = \epsilon. \label{a6}
\end{equation}
Solving \textbf{Eqn} (\ref{a6}) for \(\dot{r}^2\), we obtain the radial equation
\begin{equation}
\dot{r}^2 = E^2 + \frac{(2Mr - \frac{\alpha M^2}{r^2})(aE + L)^2}{r^4} + \frac{1}{r^2} \left( a^2 E^2 - L^2 \right) + \epsilon  \frac{\Delta}{r^2}. \label{a9}
\end{equation}

This framework enables an examination of this equation to understand its primary characteristics, including stable orbits, turning points, and the impact of quantum corrections on particle trajectories and energy extraction mechanisms. It also enables direct comparison with classical black holes by observing the effects introduced by quantum corrections.

\subsection{Angular Velocity of a Particle}

In order to determine the motion of a particle around a rotating black hole, we need to look at the angular velocity of the particle, which measures the rate at which the particle orbits the black hole as perceived by a far observer. The black hole rotation in the RQCBH geometry causes frame dragging that makes the nearby particles co-rotate with the spacetime, especially in the ergoregion \cite{bardeen1972rotating}.

The angular velocity is defined by the rate of change of the azimuthal coordinate $\phi$ with respect to coordinate time $t$
\begin{equation}
\Omega^{\text{RQCBH}} = \frac{d\phi}{dt}. \label{a10}
\end{equation}

In the ergosphere, where effects of frame-dragging are dominant, particles cannot be at rest relative to infinity. Instead, their movement is limited by the geometry to be within a permitted range of angular velocities. By imposing the condition $ds^2 \geq 0$ for timelike motion, the allowed interval is given by
\begin{equation}
\Omega^{\text{RQCBH}}_{\pm} = \frac{-g_{t\phi} \pm \sqrt{g_{t\phi}^2 - g_{tt} g_{\phi\phi}}}{g_{\phi\phi}}. \label{a11}
\end{equation}

At the static limit (where $g_{tt} = 0$), $\Omega^{\text{RQCBH}}_+ = 0$, while inside the ergosphere, both $\Omega^{\text{RQCBH}}_+$ and $\Omega^{\text{RQCBH}}_-$ become positive. This implies that all particles, regardless of their initial angular momentum, are dragged in the direction of the black hole’s rotation \cite{wang2019shadows}.

For the RQCBH spacetime, substituting the metric components of (\ref{a1}) into the expression (\ref{a11}), the angular velocity takes the form
\begin{equation}
\Omega^{\text{RQCBH}}_{\pm} = \omega^{\text{RQCBH}} \pm  
\frac{r^2 \sqrt{\Delta}}{(a^2 + r^2)^2 - a^2 \Delta},
\end{equation}
where the frame-dragging term is defined as
\begin{equation}
\omega^{\text{RQCBH}} = \frac{a \left(2 M r - \frac{\alpha M^2}{r^2} \right)}{(a^2 + r^2)^2 - a^2 \Delta} . 
\end{equation}

Taking the near-horizon limit, the particle's angular velocity converges to that of the black hole
\begin{equation}
\lim_{r \to r_+} \Omega^{\text{RQCBH}}_+ = \lim_{r \to r_+} \Omega^{\text{RQCBH}}_- = \omega^{\text{RQCBH}}_{\text{BH}} = \frac{a(2Mr_+ - \frac{\alpha M^2}{r^2})}{r_+^4 + a^2 r_+^2 - a^2 (2Mr_+ - \frac{\alpha M^2}{r^2})}.
\end{equation}

This asymptotic behavior confirms that particles near the event horizon are inevitably dragged by the black hole’s rotation. Even those with zero angular momentum at infinity are compelled to co-rotate near the horizon. In the equatorial plane of the RQCBH spacetime, the angular velocity of a particle, defined in (\ref{a10}), is given by
\begin{equation}
\Omega^{\text{RQCBH}} = \frac{d\phi}{dt} = \frac{a(2Mr^3-\alpha M^2 )E-(r^4 - 2Mr^3 + \alpha M^2)L  }{\left[r^4(r^2 + a^2) + a^2(2Mr^3 +\alpha M^2)\right]E +aL(2Mr^3 - \alpha M^2)}. \label{a15}
\end{equation}

For particles in circular equatorial orbits, the conserved energy $E$ and angular momentum $L$ are given by:
\begin{equation}
E =\frac{r^4 - 2 M r^3 + \alpha M^2 \pm a r^2 \sqrt{M r - \frac{\alpha M^2}{r^2}}}
{r^3 \sqrt{r^4 - 3 M r^3 + 2 \alpha M^2 \pm 2 a r^2 \sqrt{M r - \frac{\alpha M^2}{r^2}}}},\ \label{a16}
\end{equation}
\begin{equation}
L = \pm \frac{r^4 + a^2 r^2 \mp 2a r^2 \sqrt{M r - \frac{\alpha M^2}{r^2}} \mp a \alpha M^2}
{r^2 \sqrt{r^4 - 3 M r^3 + 2 \alpha M^2 \pm 2a r^2 \sqrt{M r - \frac{\alpha M^2}{r^2}}}}. \label{a17}
\end{equation}

Substituting \textbf{Eqs} (\ref{a16}) and (\ref{a17}) into (\ref{a15}) into the angular velocity expression yields:
\begin{equation}
\Omega^{\text{RQCBH}} = \frac{\mp r \sqrt{M r^3 - \alpha M^2}}{r^3 \mp a \sqrt{M r^3 - \alpha M^2}}.
\end{equation}

This result shows how the angular velocity depends on the black hole’s spin $a$, the deformation parameter $\alpha$, and the radial coordinate $r$. The upper and lower signs correspond to co-rotating and counter-rotating orbits, respectively.

\begin{figure}[htbp]
    \centering

    \makebox[\textwidth][c]{
        \includegraphics[width=0.4\textwidth]{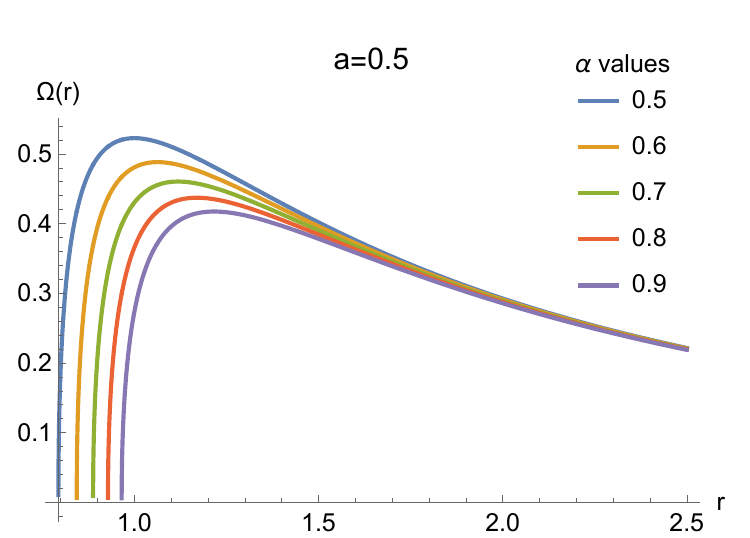}
        \hspace{2em}
        \includegraphics[width=0.4\textwidth]{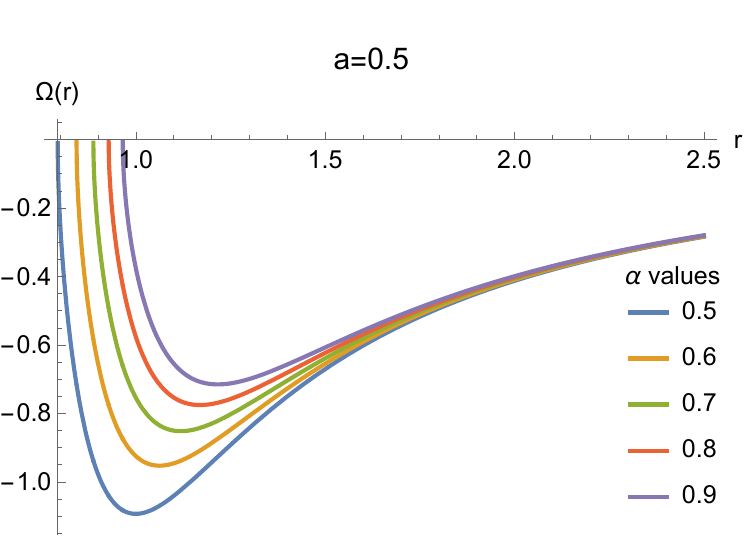}
    }

    \caption{Dependence of angular velocity $\Omega$ on $\alpha$ versus radial distance $r$.}
    \label{fig:5}
\end{figure}

The two plots in \textbf{Fig.} [\ref{fig:5}] show how the angular velocity ($\Omega$) changes with the radial coordinate $r$ for a fixed spin parameter $a$ and increasing values of the correction parameter $\alpha$. Both the co-rotating (positive $\Omega$) and counter-rotating (negative $\Omega$) branches are plotted. The results indicate that as $\alpha$ increases, the magnitude of the angular velocity decreases. This suggests that quantum corrections suppress the frame-dragging effects within the ergosphere, a phenomenon similar to the effect of positive tidal charge in braneworld models \cite{khan2019particle}.

\section{Collisional Energy Extraction in the Quantum-Corrected Ergoregion} \label{s4}

The rotating quantum-corrected black hole exhibits significant modifications in the structure and extent of the ergoregion due to quantum gravitational effects. These alterations influence the behaviour of particles near the black hole, affecting conserved quantities like energy and angular momentum. As a result, the standard conditions for energy extraction via the Penrose process are modified. In this work, we analyse how these quantum corrections give rise to new negative energy states during particle collisions, potentially enhancing the overall extraction efficiency.

\subsection{Negative Energy States}

Negative energy states play a crucial role in the Penrose process and can occur as a result of counter-rotating motion or electromagnetic interactions \cite{penrose1969gravitational, bardeen1972rotating}. Negative energy trajectories arise in the ergoregion of rotating black holes, such as the braneworld Kerr black hole, because of the strange geometry of spacetime \cite{khan2019particle}.

To analyze the energy conditions, we start with the radial \textbf{Eqn.} (\ref{a9}) of motion, which can be reformulated to examine energy bounds:
\begin{equation}
\begin{split}
& E^2\left[(r^2 + a^2)r^2 + a^2\left(2Mr - \frac{\alpha M^2}{r^2}\right)\right]
+ 2aEL\left(2Mr - \frac{\alpha M^2}{r^2}\right)\\
&- L^2\left(r^2 - 2Mr + \frac{\alpha M^2}{r^2}\right)
+ \epsilon r^2 \Delta = 0, \label{a19}
\end{split}
\end{equation}
where \( \epsilon = -1, 0, 1 \) corresponds to timelike, null, and spacelike geodesics respectively, and \( \Delta = r^2 - 2Mr + a^2 +  \frac{\alpha M^2}{r^2}\).

Solving expression (\ref{a19}) for \( E \) and \( L \), we obtain

\begin{align}
E &= \frac{ -a L \left( 2 M r - \frac{\alpha M^2}{r^2} \right) \pm Z_1 \sqrt{\Delta} }{ r^4 + a^2 \left( r^2 + 2 M r - \frac{\alpha M^2}{r^2} \right) },
\, \\
L &= \frac{-aE(2Mr - \frac{\alpha M^2}{r^2}) \pm \sqrt{Z_2 \Delta}}{-(r^2 - 2Mr + \frac{\alpha M^2}{r^2})},
\end{align}

\begin{figure}[htbp]
    \centering

    \makebox[\textwidth][c]{
        \includegraphics[width=0.45\textwidth]{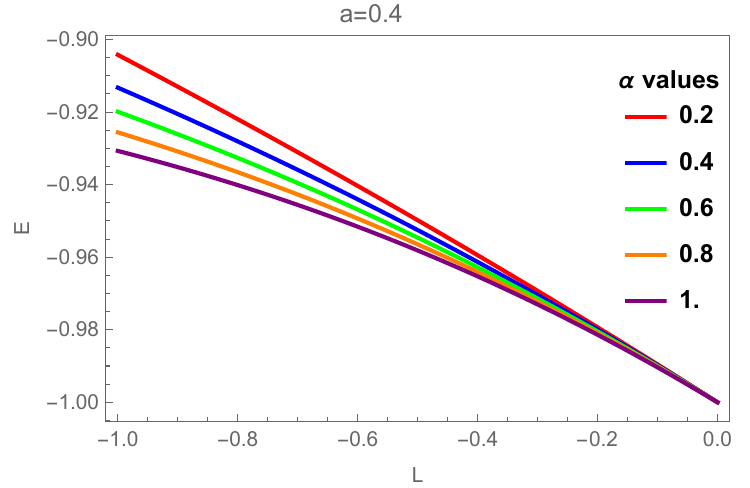}
        \hspace{2em}
        \includegraphics[width=0.45\textwidth]{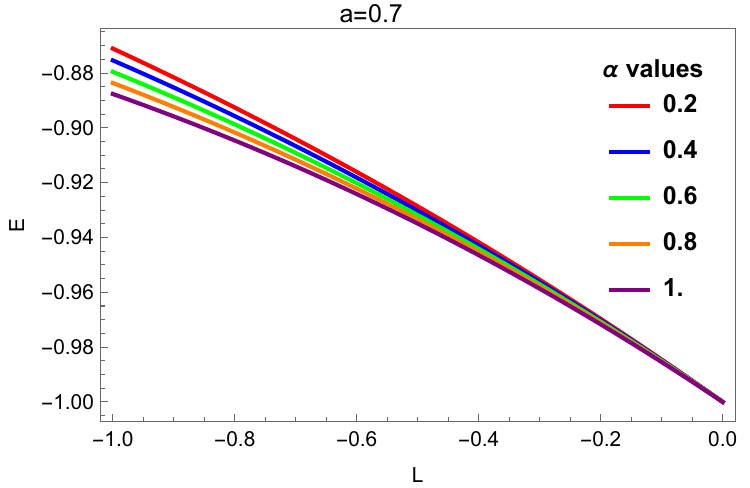}
    }

    \vspace{2em}

    \makebox[\textwidth][c]{
        \includegraphics[width=0.45\textwidth]{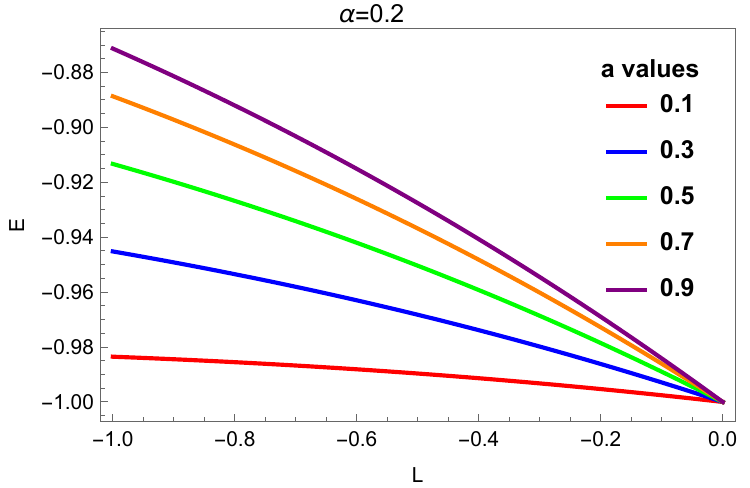}
        \hspace{2em}
        \includegraphics[width=0.45\textwidth]{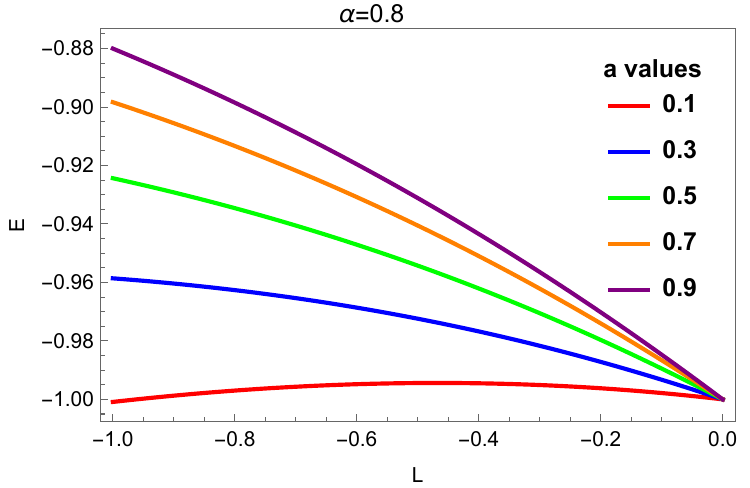}
    }

    \caption{Behavior of E versus L for negative energy states.}
    \label{fig:6}
\end{figure}

where
\begin{equation*}
Z_1 = \sqrt{
\frac{1}{r^2} \left[
L^2 r^6 - \epsilon \left( r^6 + a^2 \left( r^4 + 2 M r^3 - \alpha M^2 \right) \right)
\right]
},
\end{equation*}

\begin{equation*}
Z_2 = \sqrt{
r^4 E^2 + \epsilon \left( r^4 - 2 M r^3 + \alpha M^2 \right)
}.
\end{equation*}

The results are obtained using the following relation

\begin{equation}
r^8 \Delta - a^2 \left(2 M r^3 - \alpha M^2 \right)^2 =
r^2 \left[ r^6 + a^2 \left( r^4 + 2 M r^3 - \alpha M^2 \right) \right]
\left( r^4 - 2 M r^3 + \alpha M^2 \right). \label{a22}
\end{equation}

To identify conditions that permit negative energy (\( E < 0 \)), we fix \( E = 1 \) and examine when the expression becomes negative. It is evident that negative energy requires \( L < 0 \), and further conditions can be extracted by simplifying the inequality:
\begin{equation}
a^2 L^2 \left( 2 M r^3 - \alpha M^2 \right)^2 >
r^2 \Delta \left[ r^4 L^2 - \epsilon \left( r^6 + a^2 \left( r^4 + 2 M r^3 - \alpha M^2 \right) \right) \right]. \label{a23}
\end{equation}
By using \textbf{Eqn.} (\ref{a22}), expression (\ref{a23}) takes the form
\begin{equation}
\frac{1}{r^4}
\left( r^6 + a^2 \left( r^4 + 2 M r^3 - \alpha M^2 \right) \right)
\left[ \left( r^4 - 2 M r^3 + \alpha M^2 \right) L^2 - \epsilon r^4 \Delta \right] < 0 ,\label{a24}
\end{equation}

From inequality (\ref{a24}), it follows that
\begin{equation}
E < 0 \iff L < 0 \quad \text{and} \quad \left( \frac{r^4 - 2Mr^3 + \alpha M^2}{r^4} \right) < \frac{\epsilon \Delta}{L^2}. \label{a25}
\end{equation}

If we take $\alpha=0$, the inequality (\ref{a25}) reduces to the case of Kerr BH \cite{chandrasekhar1985mathematical}. Also, the relation shows that for a particle to attain negative energy, its angular momentum must also be negative, and its position within the ergoregion must satisfy the above inequality.

\textbf{Fig.} [\ref{fig:6}] illustrates how the particle energy $E$ varies with angular momentum $L$ in the background of RQCBH, highlighting the effects of both the spin parameter $a$ and the quantum correction parameter $\alpha$. Plots of negative energy versus angular momentum $L$ clearly show that an increasing correction parameter $\alpha$ leads to more negative energy values and a broader range of angular momenta supporting these states. Conversely, when $\alpha$ is fixed, an increase in the spin parameter $a$ makes the negative energy values less negative.

\section{Wald Inequality} \label{s5}

The Wald inequality is a crucial tool in the study of the bounds on extracting energy from a rotating black hole through the Penrose process. The fragments of a disintegrating particle within the ergoregion may have varying amounts of energy. To extract energy, at least one fragment must fly off to infinity with more energy than the initial particle \cite{hejda2019energy,parthasarathy1986high}. The Wald inequality \cite{wald1974energy} provides a theoretical limit on the amount of energy that such fragments can possess, in terms of the geometry of spacetime.
Here we obtain and use the Wald inequality for the RQCBH black hole, where the correction parameter $\alpha$ is introduced to the standard Kerr geometry.

Let a parent particle with four-velocity \( U^\mu \) and specific energy \( E = -\xi_\mu U^\mu \) decay into two fragments, one of which has four-velocity \( u^\mu \) and specific energy \( \epsilon = -\xi_\mu u^\mu \), where \( \xi^\mu = \partial_t \) is the timelike Killing vector of the spacetime.

We define an orthonormal tetrad comoving with the parent particle:
\begin{equation}
U^\mu = e^\mu_{(0)}, \quad u^\mu = \rho \left( U^\mu + v^{(i)} e^\mu_{(i)} \right), \quad \rho = \frac{1}{\sqrt{1 - v^2}}.
\end{equation}

Then the specific energy of the fragment becomes:
\begin{equation}
\epsilon = -\xi_\mu u^\mu = \rho \left( E + v^{(i)} \xi_{(i)} \right),
\end{equation}
where \( \xi_{(i)} = \xi_\mu e^\mu_{(i)} \) are the spatial components of the Killing vector in the tetrad frame.

Introducing the angle \( \theta \) between the spatial velocity \( \vec{v} \) and the spatial projection \( \vec{\xi} \), we can write:
\begin{equation}
\epsilon = \rho \left( E + |\vec{v}||\vec{\xi}|\cos\theta \right).
\end{equation}

The magnitude of the spatial Killing vector is:
\begin{equation}
|\vec{\xi}| = \sqrt{E^2 + g_{tt}},
\end{equation}
so the Wald inequality becomes \cite{ullah2019particle, liu2018energy}:
\begin{equation}
\rho E - \rho |\vec{v}| \sqrt{E^2 + g_{tt}} \leq \epsilon \leq \rho E + \rho |\vec{v}| \sqrt{E^2 + g_{tt}}.
\end{equation}

In our quantum-corrected rotating black hole model (\ref{a1}), the metric component \( g_{tt} \) is given by:
\begin{equation}
g_{tt}(r) = -\left(1 - \frac{2M}{r} + \frac{\alpha M^2}{r^4} \right).
\end{equation}

We work in geometrized units with \( M = 1 \), so:
\begin{equation}
g_{tt}(r) = -\left(1 - \frac{2}{r} + \frac{\alpha}{r^4} \right). \label{a32}
\end{equation}

We numerically searched for extremal or near-extremal configurations using the condition that the difference \( \delta r = r_+ - r_- \) between the outer and inner horizons becomes minimal. For the parameter values \( a = 0.66249 \) and \( \alpha = 0.8 \), we obtained the smallest such difference \( \delta r \approx 0.00363 \), with the outer horizon located at \( r_+ \approx 1.33752 \).

Substituting these values into  (\ref{a32})
\begin{equation}
g_{tt}(r_+) \approx -\left(1 - \frac{2}{1.33752} + \frac{0.8}{1.33752^4} \right) \approx 0.2453.
\end{equation}

For a fragment with specific energy \( E = 1 \), the Wald condition for negative energy yields
\begin{equation}
|\vec{v}| > \frac{1}{\sqrt{1 + g_{tt}(r_+)}} \approx \frac{1}{\sqrt{1.2453}} \approx 0.896.
\end{equation}

This implies that the fragment must be moving at more than 89.6\% the speed of light (relative to the parent particle) to have negative energy and enable energy extraction \cite{wagh1985revival}.

\subsection{Efficiency of Energy Extraction}

The energy extraction efficiency from a black hole via the Penrose process is a fundamental aspect in black hole energetics \cite{penrose1971extraction}. Consider a particle with energy \( E^{(0)} \) entering the ergoregion and decaying into two fragments. One of these fragments (particle 1) escapes to infinity with energy \( E^{(1)} \), while the other (particle 2) falls into the black hole with energy \( E^{(2)} < 0 \). Conservation of energy implies:
\begin{equation}
E^{(0)} = E^{(1)} + E^{(2)},
\end{equation}
which ensures \( E^{(1)} > E^{(0)} \), signifying energy extraction \cite{penrose1971extraction,parthasarathy1986high}.

Let the radial velocity of the particle relative to an observer at infinity be \( \nu = \frac{dr}{dt} \). The conserved energy and angular momentum can be expressed as:
\begin{equation}
L = p_t \, \Omega, \quad E = -p_t \, Y, \label{a36}
\end{equation}
where \( Y = g_{tt} + g_{t\phi} \, \Omega \) and \( \Omega = \frac{d\phi}{dt} \) is the angular velocity \cite{chandrasekhar1998mathematical}.

Using the normalization condition for the four-velocity, \( p^\mu p_\mu = -m^2 \), one obtains:
\begin{equation}
g_{tt} \dot{t}^2 + 2 g_{t\phi} \dot{t} \dot{\phi} + g_{rr} \dot{r}^2 + g_{\phi\phi} \dot{\phi}^2 = -m^2. \label{a37}
\end{equation}
Dividing \textbf{Eqn.} (\ref{a37}) by \( \dot{t}^2 \) and substituting \( \Omega \), we get:
\begin{equation}
g_{tt} + 2 g_{t\phi} \Omega + g_{\phi\phi} \Omega^2 + \frac{r^2 \nu^2}{\Delta} = -\left( \frac{m Y}{E} \right)^2. \label{a38}
\end{equation}

The last term on the left of \textbf{Eqn.} (\ref{a38}) is always non-negative, hence
\begin{equation}
g_{tt} + 2 g_{t\phi} \Omega + g_{\phi\phi} \Omega^2 \leq -\left( \frac{m Y}{E} \right)^2 - \frac{r^2 \nu^2}{\Delta}.
\end{equation}

Using \textbf{Eqn.} (\ref{a36}), the conservation of energy and angular momentum during the decay gives
\begin{align}
p_t^{(0)} Y^{(0)} &= p_t^{(1)} Y^{(1)} + p_t^{(2)} Y^{(2)}, \label{40} \\
p_t^{(0)} \Omega^{(0)} &= p_t^{(1)} \Omega^{(1)} + p_t^{(2)} \Omega^{(2)}. \label{41}
\end{align}

The efficiency \( \eta \) of energy extraction is defined as
\begin{equation}
\eta = \frac{E^{(1)} - E^{(0)}}{E^{(0)}} = \chi - 1, \quad \text{where} \quad \chi = \frac{E^{(1)}}{E^{(0)}}, \; \chi > 1, \label{a42}
\end{equation}

Using the \textbf{Eqns}. (\ref{a36}), (\ref{40}) and (\ref{41}), the ratio \( \chi \) in (\ref{a42}) becomes
\begin{equation}
\chi = \frac{(\Omega^{(0)} - \Omega^{(2)}) Y^{(1)}}{(\Omega^{(1)} - \Omega^{(2)}) Y^{(0)}}. \label{a43}
\end{equation}

Assuming the incoming particle with \( E^{(0)} = 1 \) splits into two photons (massless particles), the efficiency reaches its maximum when the escaping particle has minimum angular velocity and the infalling one has maximum angular velocity, i.e., 
\begin{equation}
\Omega^{(1)} = \Omega_+, \quad \Omega^{(2)} = \Omega_-  \label{a44}
\end{equation}

with both fragments released at zero radial velocity. Then:
\begin{equation}
Y^{(0)} = g_{tt} + \Omega^{(0)} g_{t\phi}, \quad Y^{(1)} = g_{tt} + \Omega_+ g_{t\phi}. \label{a45}
\end{equation}

The angular velocity \( \Omega^{(0)} \) can be found by writing \textbf{Eqn.} (\ref{a37}) as
\begin{equation}
(g_{t\phi}^2 + g_{\phi\phi}) \Omega^2 + 2 \Omega (1 + g_{tt}) g_{t\phi} + (1 + g_{tt}) g_{tt} = 0 ,
\end{equation}

\begin{table}[htbp]
\centering
\caption{Maximum efficiency $\eta$ (\%) for different values of $a$ and $\alpha$}
\renewcommand{\arraystretch}{1.2}
\begin{tabular}{|c|c|c|c|c|c|c|c|c|c|c|}
\hline
$\alpha$ & $a=0.2$ & $a=0.3$ & $a=0.4$ & $a=0.5$ & $a=0.6$ & $a=0.7$ & $a=0.8$ & $a=0.9$ & $a=1.0$ \\
\hline
0.0 & 0.25\% & 0.59\% & 1.08\% & 1.76\% & 2.70\% & 4.01\% & 5.90\% & 9.01\% & 20.71\% \\
0.1 & 0.26\% & 0.59\% & 1.09\% & 1.80\% & 2.76\% & 4.12\% & 6.15\% & 9.83\% & -- \\
0.2 & 0.26\% & 0.60\% & 1.11\% & 1.83\% & 2.83\% & 4.26\% & 6.48\% & 11.64\% & -- \\
0.3 & 0.27\% & 0.61\% & 1.13\% & 1.87\% & 2.91\% & 4.41\% & 6.92\% & -- & -- \\
0.4 & 0.27\% & 0.62\% & 1.15\% & 1.91\% & 2.99\% & 4.61\% & 7.67\% & -- & -- \\
0.5 & 0.27\% & 0.64\% & 1.18\% & 1.96\% & 3.09\% & 4.87\% & -- & -- & -- \\
0.6 & 0.28\% & 0.65\% & 1.21\% & 2.02\% & 3.21\% & 5.25\% & -- & -- & -- \\
0.7 & 0.28\% & 0.66\% & 1.24\% & 2.08\% & 3.37\% & 6.14\% & -- & -- & -- \\
0.8 & 0.29\% & 0.68\% & 1.27\% & 2.16\% & 3.59\% & -- & -- & -- & -- \\
0.9 & 0.30\% & 0.69\% & 1.31\% & 2.25\% & 3.97\% & -- & -- & -- & -- \\
1.0 & 0.30\% & 0.71\% & 1.36\% & 2.39\% & -- & -- & -- & -- & -- \\
\hline
\end{tabular}
\label{T2}
\end{table}

leading to:
\begin{equation}
\Omega^{(0)} = \frac{-(1 + g_{tt}) g_{t\phi} + \sqrt{(1 + g_{tt}) (g_{t\phi}^2 - g_{tt} g_{\phi\phi})}}{g_{t\phi}^2 + g_{\phi\phi}}.
\end{equation}

Substituting \textbf{Eqns}. (\ref{a44}) and (\ref{a45}) into (\ref{a43}) the expression for efficiency, we obtain:
\begin{equation}
\eta = \frac{(g_{tt} + g_{t\phi} \Omega_+)(\Omega^{(0)} - \Omega_-)}{(g_{tt} + g_{t\phi} \Omega^{(0)})(\Omega_+ - \Omega_-)} - 1.
\end{equation}

To determine the maximal efficiency \( \eta_{\text{max}} \), assume the decay happens arbitrarily close to the event horizon. In that limit:

\begin{equation}
\eta_{\text{max}} = \frac{1}{2} \left( \sqrt{ \frac{2 M r^3_+ - \alpha M^2}{r_+^4} } - 1 \right).
\end{equation}

\begin{figure}[htbp]
    \centering

    \begin{minipage}[b]{0.45\textwidth}
        \centering
      (a)  \includegraphics[width=\textwidth]{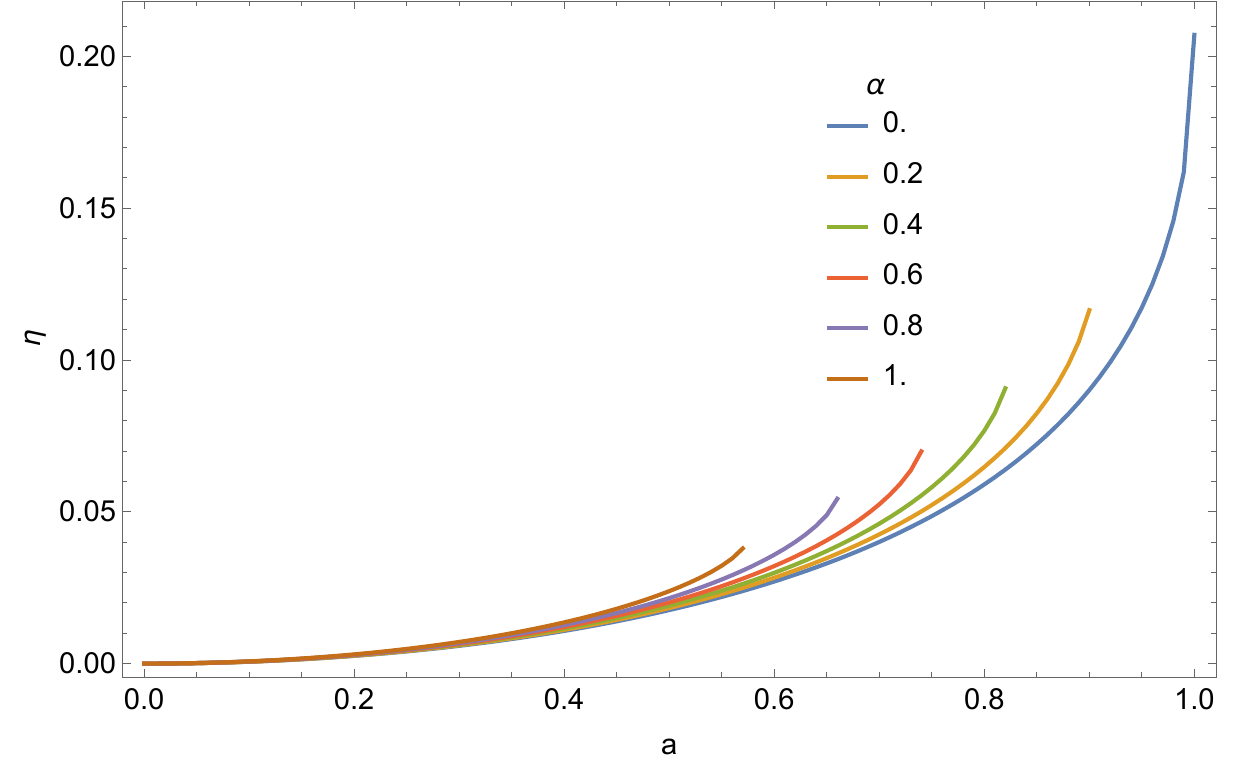}
    \end{minipage}
    \hspace{0.05\textwidth}
    \begin{minipage}[b]{0.45\textwidth}
        \centering
       (b) \includegraphics[width=\textwidth]{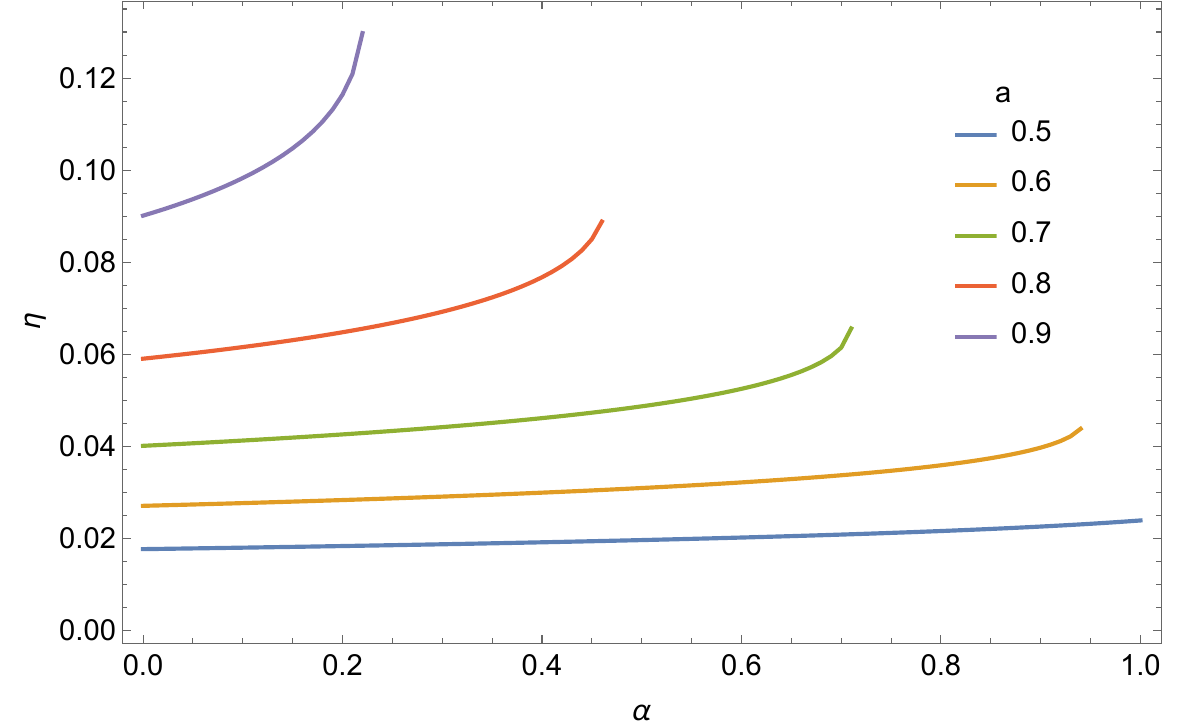}
    \end{minipage}

    \vspace{0.8cm} 

    \begin{minipage}[b]{0.45\textwidth}
        \centering
      (c)  \includegraphics[width=\textwidth]{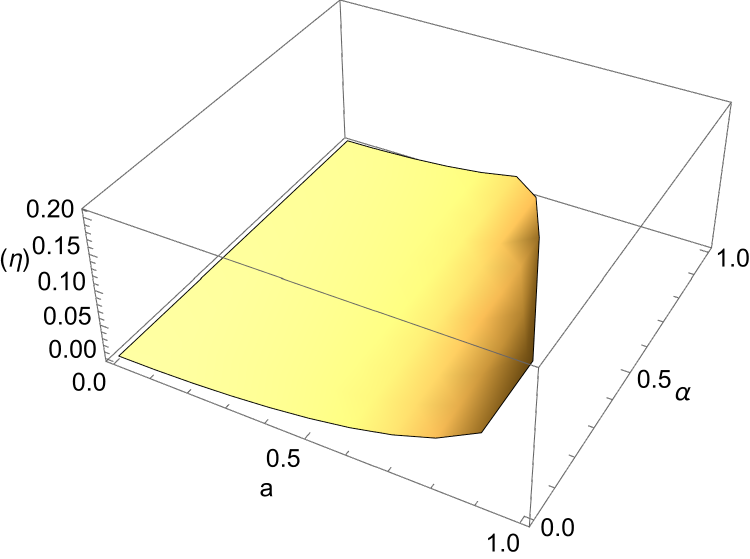}
    \end{minipage}
    
    \caption{Maximum efficiency of the energy extraction as a function a (a), function of $\alpha$ (b) and combined (c)}
    \label{fig:7}
\end{figure}

Numerical evaluations for different values of spin parameter \( a \) and quantum correction parameter  $\alpha$ in \textbf{Table} [\ref{T2}] show that the efficiency improves as both parameters increase. Notably, for the extremal Kerr black hole (\( a = 1,\ \alpha = 0 \)), the upper limit reaches \( 20.7\% \) \cite{chandrasekhar1985mathematical}, consistent with classical results. The graphical behavior in \textbf{Fig.} [\ref{fig:7}] further supports the trend that higher rotation and quantum effects enhance energy extraction capabilities.

\section{Irreducible Mass}

The irreducible mass is a fundamental quantity that sets the ultimate limit on the amount of energy that can be extracted from a rotating black hole \cite{ christodoulou1970reversible}. In the quantum-corrected rotating black hole model considered here, the irreducible mass encapsulates the geometry of the event horizon while also incorporating the effects of quantum corrections. These corrections modify the theoretical efficiency limits of energy extraction processes, such as the Penrose process \cite{carneiro2024irreducible}.

The event horizon radius, $r_+$, is determined by the largest real root of $\Delta = 0$. The irreducible mass is defined in terms of the horizon area \( A \) as
\begin{equation}
M_{\text{irr}} = \sqrt{\frac{A}{16\pi}}. \label{a50}
\end{equation}

The area of a rotating black hole is computed via the surface integral over the angular coordinates at constant \( r = r_+ \)\cite{christodoulou1970reversible}:
\
\begin{equation}
A = \int_{0}^{2\pi} d\phi \int_{0}^{\pi} d\theta \, \sqrt{g_{\theta\theta} \, g_{\phi\phi}} \bigg|_{r = r_+}, \label{a51}
\end{equation}

 where
 \begin{equation}
 \sqrt{g_{\theta\theta} \, g_{\phi\phi}} \bigg|_{r = r_+}= \sin\theta \sqrt{(r_+^2 + a^2 \cos^2\theta)^2 + 2a^2(r_+^2 + a^2 \cos^2\theta)\sin^2\theta + a^4 \sin^4\theta} ,\label{a52}
 \end{equation}

 by simplifying the expression (\ref{a52}) we get
 
 \begin{equation}
\sqrt{g_{\theta\theta} \, g_{\phi\phi}} \bigg|_{r = r_+} = \sin\theta \left( r_+^2 + a^2 \right).
\end{equation}

Therefore, the area in (\ref{a51}) becomes:
\begin{equation}
A = \int_{0}^{2\pi} \int_{0}^{\pi} (r_+^2 + a^2) \sin\theta \, d\theta \, d\phi = 4\pi (r_+^2 + a^2).
\end{equation}

The irreducible mass defined in (\ref{a50}) takes the form
\begin{equation}
    M_{\text{irr}} = \sqrt{\frac{A}{16\pi}} = \frac{1}{2} \sqrt{r_+^2 + a^2}.
\end{equation}
which generalizes the classical Kerr result by incorporating the effect of $\alpha$ through its influence on $r_+$ \cite{abbasi2025energy}.

The horizon equation determines the dependence of the $M_{\text{irr}}$ on the parameters of the black hole. With positive values of $\alpha$, the quantum correction term decreases the radius of the horizon at fixed values of $M$ and $a$, leading to a smaller irreducible mass and consequently a larger maximum extractable energy. The spin parameter $a$ still performs its classical duty, decreasing $M_{\text{irr}}$ as it grows, to the extremal limit.

\begin{table}[htbp]
\centering
\caption{Event horizon radius \(r_+\) and irreducible mass \(M_{\text{irr}}\) for different values of \(\alpha\) and spin parameter \(a\).}
\renewcommand{\arraystretch}{1.2}
\begin{tabular}{|c|c|c|c|c|c|c|}
\hline
\(\alpha\) & \multicolumn{2}{c|}{\(a = 0.2\)} & \multicolumn{2}{c|}{\(a = 0.3\)} & \multicolumn{2}{c|}{\(a = 0.5\)} \\
\hline
          & \(r_+\) & \(M_{\text{irr}}\) & \(r_+\) & \(M_{\text{irr}}\) & \(r_+\) & \(M_{\text{irr}}\) \\
\hline
0.0 & 1.9798  & 0.994936  & 1.95394 & 0.988418 & 1.86603 & 0.965926 \\
0.1 & 1.96651 & 0.988327  & 1.93991 & 0.981484 & 1.84937 & 0.957691 \\
0.2 & 1.95265 & 0.981434  & 1.92522 & 0.974229 & 1.83086 & 0.948955 \\
0.3 & 1.93816 & 0.974224  & 1.90981 & 0.966613 & 1.81153 & 0.939633 \\
0.4 & 1.92294 & 0.966658  & 1.89356 & 0.958586 & 1.79074 & 0.929613 \\
0.5 & 1.90692 & 0.958688  & 1.87635 & 0.950088 & 1.76816 & 0.918748 \\
0.6 & 1.88959 & 0.950255  & 1.85828 & 0.941042 & 1.74336 & 0.906823 \\
0.7 & 1.87191 & 0.941283  & 1.83883 & 0.931347 & 1.71568 & 0.893524 \\
0.8 & 1.85259 & 0.931676  & 1.81714 & 0.920871 & 1.68404 & 0.878350 \\
0.9 & 1.83172 & 0.921304  & 1.79394 & 0.909426 & 1.64656 & 0.860400 \\
1.0 & 1.80895 & 0.909988  & 1.76822 & 0.896745 & 1.59912 & 0.837733 \\
\hline
\end{tabular}
\label{T3}
\end{table}

To specify these dependencies more precisely, the table presents numerically computed values of the event horizon radius $r_+$, irreducible mass $M_{\text{irr}}$, and the energy extraction efficiency corresponding to different values of the quantum correction parameter $\alpha$ and spin parameter $a$. As shown in \textbf{Table} [\ref{T3}], increasing $\alpha$ leads to a monotonic decrease in both $r_+$ and $M_{\text{irr}}$, which in turn enhances the efficiency of energy extraction. This trend is consistent with the analytical expectations discussed above.

\begin{figure}[h!]
    \centering
    \includegraphics[width=0.6\textwidth]{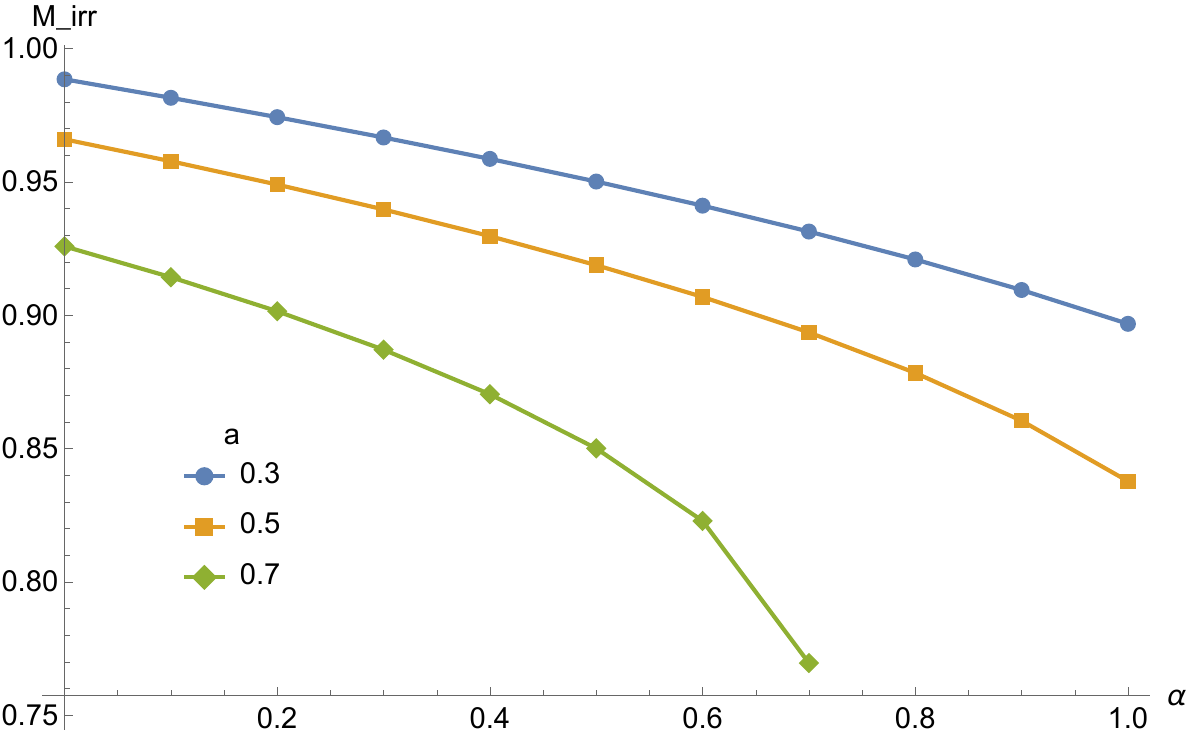}
    \caption{Variation of irreducible mass $M_{\text{irr}}$ as a function of the quantum correction parameter $\alpha$.}
    \label{fig:8}
\end{figure}

The behavior of the irreducible mass as a function of the quantum correction parameter is further visualized in \textbf{Fig} [\ref{fig:8}]. The plot of $M_{\text{irr}}$ versus $\alpha$ clearly demonstrates the sensitivity of the irreducible mass to quantum corrections, particularly for small black holes where these effects are most pronounced.

The graph confirms that quantum corrections can significantly lower the irreducible mass, thereby increasing the theoretical upper bound for extractable energy. This highlights the potential astrophysical relevance of quantum effects in the dynamics of rapidly rotating black holes.
The maximum energy that can be extracted from the black hole is given by the difference between the total mass and the irreducible mass,
\begin{equation}
    E_{\text{ext}} = M - M_{\text{irr}} ,
\end{equation}
so a smaller \( M_{\text{irr}} \) means more energy can be extracted. The correction parameter \( \alpha \) doesn't directly change the irreducible mass; instead, it influences it indirectly by modifying the event horizon radius, \( r_{+} \).

 \section{Conclusion} \label{s6}

We investigated the structural and dynamical properties of Rotating Quantum-Corrected Black Holes (RQCBH), focusing on the impact of the correction parameter $\alpha$. This study analyzed how both the black hole's spin parameter a and the quantum correction $\alpha$ influence key spacetime features, including the horizon structure, ergoregion, and static limit surfaces. Our assessment of these regions revealed the extent of deformation in the black hole geometry caused by quantum corrections and rotational effects. The results provide crucial insight into how $\alpha$ and a collectively define the causal boundaries and energy-extracting zones surrounding RQCBH.

Our results show that quantum corrections play a critical role in deforming the geometry of RQCBH. As the correction parameter $\alpha$ increases, both the event horizon and the static limit surface shift inward, effectively shrinking the black hole region. This geometric compression is accompanied by a marked expansion of the ergoregion. Analysis of the metric function \(\Delta(r)\) reveals that for sufficiently high values of $\alpha$, the spacetime approaches a non-horizon configuration, signalling a transition away from classical black hole behaviour. Furthermore, at larger values of spin and correction, the ergoregion undergoes topological changes, splitting into multiple disconnected components, highlighting the intricate sensitivity of the RQCBH structure to the combined influence of rotation and quantum corrections.

Our study explored the dynamics of test particles within the RQCBH spacetime to determine the impact of quantum corrections on orbital motion. We found that as the correction parameter $\alpha$ increases, the angular velocity of particles in the ergoregion decreases, indicating a diminished frame-dragging effect attributed to quantum phenomena. This reduced frame-dragging suggests a less efficient rotational coupling between the black hole and its immediate surroundings, subtly curtailing potential energy transfer mechanisms. Moreover, we examined the conditions necessary for negative energy states, which are essential for energy extraction, as a function of angular momentum L. Our analysis revealed that, at a fixed spin, increasing $\alpha$ leads to more profoundly negative energy states, thereby broadening the range of extractable energy. In contrast, when $\alpha$ is held constant, an increase in spin a slightly diminishes the depth of these negative energies. This outcome highlights a complex interplay between quantum corrections and rotational effects, ultimately shaping a nontrivial structure within the energy extraction phase space, even as quantum corrections generally expand the region of negative energy.

To assess the physical feasibility of energy extraction in the RQCBH spacetime, we evaluated the Wald inequality, which sets a kinematic threshold for the production of negative energy fragments. For a particle with unit energy, the condition requires that the infalling fragment must move at more than \( 89.6\% \) of the speed of light relative to the parent particle in order to acquire negative energy. This highlights a significant constraint on the Penrose process, even when geometric conditions are favorable. Despite this limitation, our analysis shows that the efficiency of the Penrose process increases with both the quantum correction parameter \( \alpha \) and the spin parameter \( a \), though the maximum efficiency attained remains modest—approaching approximately \( 11.64\% \). Additionally, we observed that the irreducible mass of the black hole decreases as \( \alpha \) increases, indicating that quantum corrections slightly lower the bound on non-extractable mass. These findings suggest that while quantum effects enhance certain aspects of the geometry and energetics, they do not fully overcome the relativistic constraints imposed by the dynamics of the process.

The persistence of this upper bound on efficiency can be attributed to several factors. Although the ergoregion enlarges and negative energy states deepen with increasing \( \alpha \), the Wald condition imposes a strict relativistic velocity threshold for fragment motion. Moreover, frame-dragging weakens slightly with increasing \( \alpha \), reducing the rotational coupling that facilitates energy extraction. The nonlinear interplay between \( \alpha \) and \( a \) also contributes, as enhancements from one parameter may be partially offset by limitations from the other. These combined effects create a natural ceiling on the extractable energy, even in a geometry that appears increasingly favourable.

In summary, our study shows that quantum corrections significantly alter the geometry and energy properties of rotating black holes. The rotating quantum-corrected black hole RQCBH features an enlarged ergoregion, deeper negative energy states, and enhanced energy extraction potential influenced by the correction parameter $\alpha$ and spin $a$. Despite these improvements, relativistic constraints like Wald’s inequality limit the process, highlighting the interplay between geometry and kinematics in energy extraction. The energy extraction efficiency of the RQCBH in this work is about $11.64\%$, lower than the $20.7\%$ efficiency of the classical Kerr black hole. In contrast, braneworld Kerr black holes can exceed Kerr efficiency depending on the brane parameter and spin, especially for negative tidal charge values.




\end{document}